\documentstyle[12pt]{article}
\begin{document}
\begin{center}
{\LARGE On the Geometry of Antisymmetric Fields}
\end{center}

\vskip1.cm \centerline{\large{M.I. Caicedo, I. Mart\'{\i}n and A. 
Restuccia}}
\vskip.5cm \centerline{\it Universidad Sim\'on Bol\'{\i}var, 
Departamento de
F\'{\i}sica} \centerline{\it Apartado Postal 89000, Caracas 1080-A,
Venezuela} \vskip.5cm \centerline{email: mcaicedo@fis.usb.ve,
isbeliam@usb.ve, arestu@usb.ve}
\vskip1.cm 

\noindent{\bf Abstract} \vskip.5cm

\begin{quotation}
{\small 
\noindent{}The geometry of antisymmetric fields with nontrivial 
transitions over
a base manifold
is described in terms of exact sequences of cohomology groups. This 
formulation
leads naturally to the appereance of nontrivial topological charges 
associated to 
the periods of the curvature of the antisymmetric fields. The relation 
between the
partition functions of dual theories is carefully studied under the 
most general assumptions,
and new topological factors related to zero modes and the Ray Singer 
torsion are found}
\end{quotation}

\vskip.5cm

Antisymmetric fields with nontrivial topological charges play a 
relevant
role in the dynamics of $p$-branes and Dirichlet brane theories, and 
more
generally in the dynamics of M-theory.Yet a few is known about the
geometrical aspects related to them. Most of the analysis in the 
literature
has been performed in terms of the associated curvatures. However, the
corresponding quantum field theory requires a formulation in terms of
antisymmetric fields with non trivial transitions over the base 
manifold, in
the same way as quantum electrodynamics needs a description in terms 
of 1-form connections
while their classical field equations may be expressed in terms of the 
curvature only.

For line bundles with connections a theory exists  \cite{re:Weil} for
a globally defined closed 2-form $\Omega $ on a manifold that may be
summarized as follows: $\Omega $ is the curvature of a connection over 
some
line bundle if and only if its cohomology class divided by $2\pi {}i$ 
is
integral. A generalization of this theory to 3-forms has been 
developed
in \cite{re:Brylinski}. Also, some more recent work in 
\cite{re:Murray}. 
Brylinski's  approach in \cite{re:Brylinski} is in terms of 
characteristic
classes of gerbes. Gerbes are fiber bundles whose fibers are grupoids. 
Its
generalization to 4-forms may be obtained \cite{re:private} but there
is not a general approach for $p$-forms. Even when a geometrical 
structure 
for  $p$-forms equivalent to gerbes is not known, a description
in terms of local $p$-forms over a manifold, with transitions 
satisfying
higher order cocycle conditions may be implemented rigourously. In 
quantum field theories,
the latter is sufficient for the construction of models of 
antisymmetric tensor field theories
with a complete topological description in terms of exact sequences of
cohomological groups.

In the current literature, antisymmetric tensors fields are described 
in terms of  global $
(p-1)$-forms defined over a manifold $X$, i.e forms with trivial 
transition 
functions.  However,  to describe antisymmetric 
fields in the most general way, we have to take into account fields 
with nontrivial 
transitions over the base manifold with their ''curvature'' being a 
globally 
defined $p$-form. These field configurations are the ones 
responsible for the appearence of topological charges.This latter 
point is
essential in the description of $p$-branes and $D$-branes 
from a quantum field theory point of view.

The construction of the above mentioned topological structure in terms 
of
antisymmetric tensor fields and their transitions satisfying cocycle
conditions of higher order was considered in \cite{re:CMR}, see also 
\cite
{re:Freed}. In this paper we continue along these lines and present 
some new
results concerning the topological factors appearing in the partition
function of the action describing a $(p-1)$-form with nontrivial 
transitions
over the base manifold. Explicit examples of antisymmetric fields with
nontrivial transitions over $S^{2}$ were presented in 
\cite{re:Nepomichi}
The solutions presented lead to the Dirac monopole solutions over 
$S^{2}$.
These results are particular cases of the known isomorphisms of 
\v{C}ech
cohomology groups

\[
\check{H}^{2}(S^{2},Z)\cong \check{H}^{3}(S^{3},Z) 
\]

\noindent{}the left hand side being the Chern characteristic classes
classifying the Dirac monopoles over $S^{2}$.

To begin the discussion, we will briefly review the field theory 
formulation
of $1$-form  connections with nontrivial transitions (nontrivial line
bundles) and then consider the generalization to antisymmetric fields.

The isomorphism classes of line bundles with connections may be 
represented
by equivalence classes of doublets ($A,g$) where $A$ is a $1$-form 
connection
over the base manifold $X$ and $g$ are the transition functions with
values in $C^{*}$ -the nonzero complex numbers-. $X$ is assumed to be 
an
euclidean, orientable $d$ dimensional compact manifold without 
boundary. To
describe the transition functions, an open covering $\{U_{i},i\in 
{}I\}$ of $
X $ is given. All the results expressed in terms of cohomology classes 
are,
of course, independent of the covering used. Without loosing 
generality, it
may be assumed that all the open sets $U_{i}$ of the covering are 
contractible to a point.
For any $U_{i}$, one
has a local $1$-form $A_{i}$, and on the intersections $U_{i}\cap
{}U_{j}\neq {}\emptyset $

\begin{equation}
A_{i}-A_{j}=g_{ij}^{-1}dg_{ij}  \label{eq:doublet5}
\end{equation}

\noindent {}where $g_{ij}$ are the transition functions on $U_{i}\cap
{}U_{j} $. The transition functions are required to satisfy the $2$ 
cocycle
condition {}on the intersection of any three open sets ($U_{i}\cap
{}U_{j}\cap {}U_{k}$)

\begin{equation}
(\delta {}g)_{ijk}=g_{ij}g_{jk}g_{ki}=I\mbox{.}  \label{eq:doublet6}
\end{equation}
in terms of $\ g_{ij}=e^{i\Lambda _{ij}}$this condition becomes

\begin{equation}
\Lambda _{ij}+\Lambda _{ij}+\Lambda _{ij}=2\pi n,\mbox{ }n\in Z
\label{eq:calib3}
\end{equation}

Two doublets are equivalent (gauge equivalent in physical language)

\begin{equation}
(A^{\prime},g^{\prime})\sim{}(A,g)  \label{eq:doublet1}
\end{equation}

\noindent{}if and and only if, for any $U_i$ there is a map

\begin{equation}
h_i:U_i\rightarrow{}C^{*}  \label{eq:doublet2}
\end{equation}

\noindent{}such that

\begin{equation}
A_{i}^{\prime }=A_{i}+h_{i}^{-1}dh_{i}  \label{eq:doublet3}
\end{equation}
and on $U_{i}\cap {}U_{j}$

\begin{equation}
g_{ij}^{\prime }=h_{i}g_{ij}h_{j}^{-1}.  \label{eq:doublet4}
\end{equation}

In terms of $h_{i}=e^{i\Lambda _{i}}$ and $g_{ij}=e^{i\Lambda _{ij}}$ 
the
gauge transformations (\ref{eq:doublet3})(\ref{eq:doublet4}) are 
usually
written as:

\begin{equation}
A_{i}^{\prime }-A_{i}=d\Lambda _{i}  \label{eq:calib1}
\end{equation}
\begin{equation}
\Lambda _{ij}^{\prime }=\Lambda _{ij}+\Lambda _{i}-\Lambda _{j}
\label{eq:calib2}
\end{equation}

\noindent {}The cocycle condition (\ref{eq:doublet6}) is expressed in 
terms
of the coboundary operator $\delta $ introduced by \v{C}ech to define 
the \v
{C}ech complexes and the corresponding \v{C}ech cohomology groups 
\cite
{re:Steenrod}. We should remark that between the \v {C}ech cohomology 
groups
(denoted by $\check{H}^{p}(X,\mbox{ })$) the groups 
$\check{H}^{p}(X,\Re )$
with coefficients on the constant sheaf $\Re $ (real) are isomorphic 
to the
de Rham cohomology groups ($H_{DR}^{p}(X)$), i.e.

\begin{equation}
\check{H}^p(X,\Re{})\cong{}H_{DR}(X)  \label{eq:ChechRam}
\end{equation}

There is also the following isomorphism between \v{C}ech cohomology 
groups:

\begin{equation}
\check{H}^{1}(X,\underline{C}_{X}^{*})\cong {}\check{H}^{2}(X,Z)
\label{eq:ChechChech}
\end{equation}

\noindent {}where $\underline{C}_{X}^{*}$ is the sheaf of smooth 
functions
over $X$ with values on $C^{*}$ and $Z$ is the sheaf with values on 
the
integers. $\check{H}^{1}(X,\underline{C}_{X}^{*})$ may be identified 
with
the group of topological line bundles over $X$, while 
$\check{H}^{2}(X,Z)$
is the \v{C}ech cohomology group associated to the characteristic 
Chern classes 
of line bundles.

The group of isomorphism classes of bundles with connection may be
identified with the degree $1$ \v {C}ech hypercohomology group 
$\check{H}%
^{1}(X,K)$ of the complex $K^{1}$

\begin{equation}
K^{1}:\underline{C}_{X}^{*}\stackrel{dlog}{\rightarrow }A_{X,C}^{1}
\label{eq:FibBun1}
\end{equation}

\noindent {}where $A_{X,C}^{1}$ is the sheaf with values on the 
complex $1$%
-forms over $X$.

$\check{H}^1(X,K^1)$ is canonically isomorphic to

\begin{equation}
Z(-1)\otimes{}H^{2}(X,Z(2)^{\infty}_{d})  \label{eq:FibBun2}
\end{equation}

\noindent {}where $H^{2}(X,Z(2)_{d}^{\infty })$ is the smooth Deligne
cohomology group of degree $2$. The relation between 
(\ref{eq:ChechChech})
and (\ref{eq:FibBun2}) is shown in the following exact sequence of
cohomology groups,

\begin{equation}
0\hookrightarrow {}A_{X}^{1}/A_{X,0}^{1}\rightarrow
{}H^{2}(X,Z(2)_{d}^{\infty })\rightarrow 
\check{H}^{2}(X,Z(2))\rightarrow {}0
\label{eq:FibBun3}
\end{equation}

\noindent {}where $A_{X}^{1}$ is the group of $1$-forms over $X$ and $
A_{X,0}^{1}$ is the group of closed $1$-forms with integral periods. 
This
exact sequence shows that if we take a line bundle with  $1$-form 
connection $
A$, adding to $A$ an element $\theta $ of $A_{X}^{1}/A_{X,0}^{1}$ we
obtain another line bundle with connection which is mapped to the same 
element of $
\check{H}^{2}(X,Z(2))$ i.e. the line bundle with connection $A$ and 
the one
with connection $A+\theta $ are of course on the same Chern class.
Conversely, given a line bundle with two 1-form connections on it, 
they
differ by an element of $A_{X}^{1}/A_{X,0}^{1}$.

The other relevant sequence which in particular contains Weil's 
theorem \cite{re:Weil} 
referred to before is:

\begin{equation}
0\hookrightarrow {}\check{H}^{1}(X,C^{*})\rightarrow {}\check{H}
^{2}(X,K^{1})\rightarrow {}A_{X,closed}^{2}\rightarrow 
\check{H}^{2}(X,C^{*})
\label{eq:FibBun4}
\end{equation}

\noindent here $A_{X,closed}^{p}$ is the group of closed $p$-forms 
over $X$
, and $C^{*}$ stands for the constant sheaf with values on the nonzero
complex numbers. This sequence tells us that to each topological line 
bundle
there is an associated closed $2$-form with integer periods and 
conversely:
given a closed $2$-form with integer periods there exists a line 
bundle and
a connection on it whose curvature is the given $2$ form. Moreover the
sequence (\ref{eq:FibBun4}) tells us that $\check{H}^{2}(X,K)$ is an
extension of the group of closed $2$-forms with integer periods by the 
group 
$\check{H}^{1}(X,C^{*})$, the latest being the group of line bundles 
with
constant transition functions. Over any such line bundle there exists 
a flat
$1$-form connection , so the group $\check{H}^{1}(X,C^{*})$ also 
classifies
the flat $1$-form connections over $X$ modulo 
$A_{X,closed}^{1}/A_{X,0}^{1}$
. In \cite{re:CMR} a generalization of all this construction to higher 
order 
$p$- forms was considered, the interesting point is that it makes use 
of the
generalization of (\ref{eq:FibBun1}), (\ref{eq:FibBun2}), 
(\ref{eq:FibBun3})
and (\ref{eq:FibBun4}), which are well stablished from a mathematical 
point
of view. In the latter part of this paper, we will discuss some new
topological contributions from this geometrical structure to the 
partition
function of antisymmetric tensor fields.

We will now propose a generalization of the above
results for the set of triplets $(B,\eta {},\Lambda {})$ . $B$ is a
local $2$-form, $\eta $ a local1-form and $\Lambda $ the transition
functions for the $1$-forms. The generalization to $p$-plets goes 
along
the same lines. On a covering $\{U_{i},i\in {}I\}$ of $X$ we have 
$2$-forms $B_{i}$ on $U_{i}$ which on $U_{i}\cap {}U_{j}\neq \emptyset 
$ satisfy

\begin{equation}
B_{i}-B_{j}=d\eta _{ij}  \label{eq:triplet1}
\end{equation}
$\eta _{ij\mbox{ }}$\noindent {}being 1-forms defined on $U_{i}\cap 
{}U_{j}$
, which on the intersection of three open sets i.e. when $U_{i}\cap
{}U_{j}\cap {}U_{k}\neq \emptyset $

\begin{equation}
\eta_{ij}+\eta_{jk}+\eta_{ki}=d\Lambda_{ijk}  \label{eq:triplet2}
\end{equation}

\noindent {}while on $U_{i}\cap {}U_{j}\cap {}U_{k}\cap {}U_{l}\neq
\emptyset $,  $\Lambda $ satisfies the $3$-cocycle condition 
\begin{equation}
(\delta \Lambda {})_{ijkl}=\Lambda _{ijk}-\Lambda _{ijl}+\Lambda
_{ikl}-\Lambda _{jkl}=2\pi {}n.  \label{eq:triplet3}
\end{equation}
(\ref{eq:triplet1}), (\ref{eq:triplet2}) and (\ref{eq:triplet3}) 
define a
triplet $(B,\eta {},\Lambda {})$ on a covering of $X$ (compare with 
(\ref
{eq:doublet5}),(\ref{eq:doublet6})). Two triplets $(B,\eta {},\Lambda 
{})$
and $(\tilde{B},\tilde{\eta },\tilde{\Lambda })$ are equivalent when 
they satisfy

\begin{equation}
\hat{B}_i=B_i+d\omega_i\qquad\mbox{on }U_i  \label{eq:gaugetrip1}
\end{equation}

\noindent {}for $1$-forms $\omega _{i}$ defined on $U_{i}$,

\begin{equation}
\hat{\eta}_{ij}=\eta_{ij}+\eta_i-\eta_j+d\Lambda_{ij}\qquad\mbox{on 
}U_i\cap{
}U_j\neq\emptyset  \label{eq:gaugetrip2}
\end{equation}

\noindent {}for $0-$forms $\Lambda _{ij}$ defined on $U_{i}\cap 
{}U_{j}\neq
\emptyset $

\begin{equation}
\widehat{\Lambda }_{ijk}=\Lambda _{ijk}+\Lambda _{ij}+\Lambda 
_{jk}+\Lambda
_{ki}\mbox{,
on }U_{i}\cap {}U_{j}\cap {}U_{k}\neq \emptyset  \label{eq:gaugetrip3}
\end{equation}

\noindent {}rules (\ref{eq:gaugetrip1}), (\ref{eq:gaugetrip2}) and 
(\ref
{eq:gaugetrip3}) define the gauge transformations on the space of 
triplets
and clearly generalize (\ref{eq:calib1})-(\ref{eq:calib2}).

With the above definitions the equivalence classes of triplets are
independent of the covering. These definitions lead to appropiate 
generalizations of the sequences given
in (\ref{eq:ChechChech}), (\ref{eq:FibBun1}), (\ref{eq:FibBun2}), 
(\ref
{eq:FibBun3}) and (\ref{eq:FibBun4}). Indeed, one may show that the
isomorphism $\check{H}^{1}(X,\underline{C}_{X}^{*})\cong {}\check{H}%
^{2}(X,Z) $ generalizes to the following isomorphism between 
cohomology
groups

\begin{equation}
\check{H}^{2}(X,\underline{C}_{X}^{*})\cong \check{H}^{3}(X,Z)
\end{equation}
where $\check{H}^{2}(X,\underline{C}_{X}^{*})$ is the \v {C}ech 
cohomology
group of degree $2$ identified with the transitions $\Lambda $ 
satisfying
the $3$-cocycle condition (\ref{eq:triplet3}), $\check{H}^{3}(X,Z)$ is
related but not isomorphic to the $3$-forms with integer periods as we 
will
discuss shortly, and in fact it generalizes the Chern classes of 
doublets to
triplets.

The equivalence classes of triplets may be identified with the 
\v{C}ech
hypercohomology of the complex

\begin{equation}
K^{2}:\underline{C}_{X}^{*}\stackrel{dlog}{\rightarrow }A_{X,C}^{1}%
\rightarrow {}A_{X,C}^{2}
\end{equation}

\noindent{}which will be denoted $\check{H}^2(X,K^2)$. It turns out 
that $%
\check{H}^2(X,K^2)$ is canonically isomorphic to

\begin{equation}
Z^{2}(-1)\otimes{}H^3(X,Z(3)^\infty_d)
\end{equation}

\noindent{}where $H^3(X,Z(3)^\infty_d)$ is the smooth Deligne 
cohomology of
degree $3$.

The generalization of (\ref{eq:FibBun3}) is the exact sequence

\begin{equation}
0\hookrightarrow {}A_{X}^{2}/A_{X,0}^{2}\rightarrow
{}H^{3}(X,Z(3)_{d}^{\infty })\rightarrow 
{}\check{H}^{3}(X,Z(3))\rightarrow {}0
\label{eq:alv21}
\end{equation}

\noindent while the generalization of (\ref{eq:FibBun4}) is the exact
sequence

\begin{equation}
0\hookrightarrow \check{H}^{2}(X,C^{*})\rightarrow \check{H}
^{2}(X,K^{2})\rightarrow {}A_{X,closed}^{3}\rightarrow 
\check{H}^{3}(X,C^{*})
\label{eq:exact22}
\end{equation}

\noindent Notice that to each triplet $(B,\eta {},\Lambda {})$ there 
is an
associated closed $3$-form belonging to ${}A_{X,closed}^{3}$ (the 
`curvature'
of $B$) which is then mapped to the identity element of the group 
$\check{H}%
^{3}(X,C^{*})$. Now we have also the complex of constant sheafs

\begin{equation}
0\hookrightarrow{}Z\rightarrow{}C\rightarrow{}C^{*}\rightarrow{}0
\end{equation}

\noindent{}and the corresponding large exact sequence of cohomology 
groups

\begin{equation}
\dots\rightarrow\check{H}^3(X,Z)\rightarrow\check{H}^3(X,C)\rightarrow
\check{%
H}^3(X,C^{*})\rightarrow\dots
\end{equation}

\noindent {}which shows that the identity element of 
$\check{H}^{3}(X,C^{*})$
corresponds to the image of $\check{H}^{3}(X,Z)$ onto $\check{H}%
^{3}(X,C^{*}) $ and hence, the curvature of $B$ has integral periods.
Conversely, given a closed $3$-form with integral periods the fact 
that the
map from $\check{H}^{2}(X,K^{2})$ onto ${}A_{X,closed}^{3}$ in (\ref
{eq:exact22}) is surjective, implies that there exists an equivalence 
class
of triplets for which the $3$-form is the `curvature' of the 
antisymmetric
field $B$. The correspondence between the groups of $3$-forms $L$ with
integral periods and equivalence classes of triplets is not one to 
one. In
fact, $\check{H}^{2}(X,K^{2})$ is an extension of the group 
$A_{X,0}^{3}$
(closed $3$-forms with integer periods) by the group 
$\check{H}^{2}(X,C^{*})$
the degree 2 \v {C}ech cohomology with coefficients in the constant 
sheaf $%
C^{*}$. As we stated before, all these results for $3$-plets may be
straightforwardly generalized to equivalent classes of $p$-plets. In
particular, the generalization of (\ref{eq:exact22}) given by the
following exact sequence

\begin{equation}
0\hookrightarrow \check{H}^{p}(X,C^{*})\rightarrow \check{H}%
^{p}(X,K^{p})\rightarrow {}A_{X,closed}^{p+1}\rightarrow \check{H}%
^{p+1}(X,C^{*})  \label{eq:fullexact}
\end{equation}
yields the exact relation between the group of closed $(p+1)$-forms 
with
integral periods and the equivalence classes of $(p+1)$-plets.

We will now use all these relations on the evaluation of the partition
function of the following action

\begin{equation}
S(B)=\frac{1}{2}g^{-2}\int_{X}{}dB\wedge {}^{*}dB
\end{equation}
describing a local antisymmetric tensor field $B$ (local $p$-form) 
which
satisfies the transitions of a $p$-form in an equivalence class of 
$(p+1)$-plets. The
partition function of this system (${\cal {Z}}(p,g)$) is defined by 
summing $%
e^{-\frac{1}{2g^{2}}\int_{X}dB\wedge {}^{*}dB}$ on all the equivalence
classes of $(p+1)$-plets.

We notice that from the exact sequence (\ref{eq:exact22}) the 
curvature $%
H\equiv {}dB$ must satisfy the following conditions

\begin{equation}
dH=0
\end{equation}

\begin{equation}
\oint_{\Sigma _{p}^{I}}H= 2\pi {}n^{I}
\end{equation}
where $\Sigma _{P}^{I}$ is a basis of the integer homology of 
dimension $p$
over $X$.

In order to study the partition function of $S(B)$ we now introduce 
the
following master action

\begin{equation}
S(L,V)=\frac{1}{2}g^{-2}\int_{X}(L\wedge {}^{*}L+iL\wedge {}dV)
\label{eq:masteraction}
\end{equation}

\noindent {}where $L$ is a globally defined $(p+1)$-form and $V$ a 
local 
$(d-p-2)$-form with transition given by the equivalence class of 
$(d-p-1)$-plets.

From the sequence (\ref{eq:fullexact}), it may be realized that the 
volume of
the zero mode space associated to this action is

\begin{equation}
Vol(\check{H}^{(d-p-2)}(X,\Re
{}/Z))Vol(A_{X,closed}^{(d-p-2)}/A_{X,0}^{(d-p-2)})  
\label{eq:zeromodevol}
\end{equation}

\noindent {}where $A_{X,closed}^{d-p-2}/A_{X,0}^{d-p-2}$ is the group 
of
closed $(d-p-2)$-forms modulo closed $(d-p-2)$-forms with integer 
periods, this
volume factorizes out from the partition function and, consequently, a
summation of the non zero modes still remains.

The gauge group $G_{V}$ of $S(L,V)$ comes only from the local $V$ 
form. The
functional integral on $V$\ is on the equivalence classes of 
$(d-p-1)$-plets,
and hence if we integrate on all $(d-p-1)$-plets we must divide by the 
volume
of $G_{V}$ which is defined by the generalization of 
(\ref{eq:gaugetrip1}),(
\ref{eq:gaugetrip2}) and (\ref{eq:gaugetrip3}).

Using (\ref{eq:alv21}) the integration on the classes of 
$(d-p-1)$-plets may
be performed in two steps, first one integrates on the group $
A_{X}^{d-p-2}/A_{X,0}^{d-p-2}$ for a fixed element of ${}\check{H}
^{d-p-1}(X,Z(d-p-1))$. Since the elements of 
$A_{X}^{d-p-2}/A_{X,0}^{d-p-2}$
are globally defined forms the integration on the first step reduces 
to the
standard one used in QFT. The second step consists on a generalization 
of
the summation over topological line bundles in the case of 1-form 
connection.

The previous factorization of $Vol(\check{H}^{(d-p-2)}(X,\Re {}/Z))$ 
means
that the integration on $\check{H}^{d-p-1}(X,Z(d-p-1))$ reduces to a 
sum
over the integer periods of $dV$ i.e one first fixes a period and sums 
over
all $(d-p-2)$-forms, and then sums over all periods.

We will now show that the summation technology just described leads to 
a duality
relation between ${\cal {Z}}(p,g)$ and ${\cal {Z}}
(d-p-2,g^{-1})$ containing non trivial topological factors.

Functional integration of the action (\ref{eq:masteraction}) on $L$ 
yields

\begin{equation}
g^{ b^{p+1}}{\cal {Z}}{}(d-p-2,g^{-1})  \label{eq:alv30a}
\end{equation}
\noindent {}with 
\begin{equation}
{\cal {Z}}(d-p-2,g^{-1})=\int 
{}DV\frac{1}{\mbox{Vol}G_{V}}e^{-\frac{g^{2}}{2%
}\int_{X}dV\wedge {}^{*}dV}.  \label{eq:alv30b}
\end{equation} where $ b^{p+1}$ denotes the dimension of the space of 
$(p+1)$-forms.

\noindent {}In the same way, to integrate the action 
(\ref{eq:masteraction})
on $V$, one must first
construct the BRST invariant effective action in terms of the ghosts,
antighosts and Lagrange multipliers associated to the gauge freedom of 
$V$.To do so, we will consider the following action

\begin{equation}
S(L,V,\rho )=\frac{1}{2}g^{-2}\int_{X}[(L+d\rho {})\wedge 
{}^{*}(L+d\rho
{})+iL\wedge {}dV]  \label{eq:equivaction}
\end{equation}
containing a globally defined $p$-form $\rho $, the global character 
of $
\rho $ will guarantee that this action is quantum mechanically 
equivalent to
the master action (\ref{eq:masteraction}).

The gauge symmetries of $S(L,V,\rho {})$ are

\[
V\rightarrow {}V+d\theta {}, 
\]

\[
L\rightarrow {}L+d\Lambda 
\]

\[
\rho \rightarrow {}\rho {}-\Lambda 
\]
where $\theta $ and $\Lambda $ are are globally defined forms of 
degrees $
(d-p-3)$ and $p$ respectively.

To show the quantum equivalence of (\ref{eq:equivaction}) and (\ref
{eq:masteraction}) one may consider the gauge fixing condition:

\begin{equation}
\rho{}=0
\end{equation}

\noindent {}which reduces (\ref{eq:equivaction})  to (\ref
{eq:masteraction}), and notice that the factor coming from the 
functional
integral over $\rho $ and its associated ghost, antighosts and 
Lagrange
multipliers is $1$.

We will now use (\ref{eq:equivaction}) to show the duality mentioned 
above.
We begin by imposing the following reducible gauge fixing condition on 
$L$

\begin{equation}
\chi {}= ^{*}d^{*}L  \label{eq:alv33}
\end{equation}

\noindent {}together with the ghost-for-ghost sequence. At this point 
it is
interesting to remark that all the gauge fixing and Fadeev Popov terms 
are
exactly the same ones used for a topological ''$B\wedge{}F(A)"$ theory 
since the
symmetries for $L$ are the same as those of the $B$ field while the 
symmetries
for $V$ are the same as the ones of the $A$ field. The evaluation of 
the partition
function for the BF theory \cite{re:BlauThom} was performed by 
considering a
Gaussian integration on the set of all fields. However it may also be 
done
by integrating first on all the ghost for ghost, antighost and 
Lagrange
multipliers together with the $V$ field while taking care of 
extracting the
zero mode sector of all these fields. By so doing, one is left in the 
case 
of the action (\ref{eq:equivaction}) with an integration only on the 
$L$ and $\rho$ fields,
together with a factor

\begin{equation}
\int {}D\hat{A}\frac{\delta {}(L-d\hat{A})}{\mbox{Vol}{}ZM(p)}
\label{eq:alv34}
\end{equation}

\noindent {}and the contribution of the determinants of the laplacians 
of 
$0$-, $1$-, etc. forms which combine to give a function of the Ray 
Singer
torsion \cite{re:Muller}. In the case $p=2$, $d=5$ for example, we 
obtain

\begin{equation}
T(5)^{-1}
\end{equation}
where $T(5)$ is the Ray Singer torsion of the $5$ dimensional base 
manifold $
X$. In the general case for a $(p+1)$-form $L$ on a base maniofold of
dimension $d$ we obtain

\begin{equation}
T(d)^{\alpha }
\end{equation}

\begin{equation}
\alpha =\{
\begin{array}{l}
\frac{d-2p-3}{d}(-1)^{p+1}\mbox{,  $d$ even} \\ [1.5mm] 
(-1)^{p+1}\mbox{,  $d$ odd}
\end{array}
\end{equation}

\noindent {}In (\ref{eq:alv34}),$ZM$ denote the zero mode subspace of 
the corresponding field
space. There are two contributions, one from the zero modes of $V$ 
given by (%
\ref{eq:zeromodevol}) and the other from $\widehat{A}$ .
The zero mode space for $\widehat{A}$ is

\begin{equation}
Vol(\check{H}^p(X,\Re{}/Z))Vol(A^p_{closed}/A^p_0),
\end{equation}

\noindent{}the functional integration in (\ref{eq:alv34}) is over the 
local $
p$-forms with transitions given by a $p$-plet 
$(\hat{A},\dots{},\Lambda{})$.
The $\hat{A}$ field must additionally satisfy the gauge choice 
(\ref{eq:alv33}%
), that is, $d\hat{A}$ is a harmonic $(p+1)$-form. Functional 
integration on $
L $ yields the following term in the action

\begin{equation}
\frac{1}{2}g^{-2}(d\hat{A}+ d\rho {})\wedge ^{*}(d\hat{A}+ d\rho {})
\end{equation}
now any closed $(p+1)$-form decomposes into its harmonic and exact 
parts.
Hence one may replace the term $(d\hat{A}+\rho {})$ by $dB$

\begin{equation}
dB=d\hat{A}+d\rho
\end{equation}
where $dB$ is a closed $(p+1)$-form. We end up with the formula 
\[
\frac{\mbox{Vol}ZM(d-p-2)}{\mbox{Vol}ZM(p)}T(d)^{\alpha }{\cal 
{Z}}(p,g) 
\]

where

\begin{equation}
{\cal {Z}}(p,g)=\int {}DB\frac{1}{\mbox{Vol}{}G_{B}}e^{-\frac{1}{2}%
g^{-2}\int {}dB\wedge {}^{*}dB}  \label{eq:alv39}
\end{equation}

\noindent(\ref{eq:alv39}) must be equal to (\ref{eq:alv30b}) since 
both were
deduced from the partition function of the master action (\ref
{eq:masteraction}). We then have the relation

\begin{equation}
g^{ b^{p+1}}\frac{{\cal {Z}}(d-p-2,g^{-1})}{\mbox{Vol}ZM(d-p-2)}%
=T(d)^{\alpha }\frac{{\cal {Z}}(p,g)}{\mbox{Vol}ZM(p)}  
\label{eq:alv41}
\end{equation}

\noindent{}Some words are worth adding about this last formula, the 
topological
factor $T(d)$ is a property of the base manifold, its value is $1$ for 
even $
d$ and $\neq 1$ for odd dimensions which explains why it has not been
pointed out before in the physics literature. The zero mode space 
contains
the generically nontrivial cohomology group $\check{H}(X,\Re {}/Z)$ 
with
values on the constant sheaf $\Re {}/Z$. The factor $g^{ b^{p+1}}$ has
been derived in different ways in the literature \cite{re:Witten}.

It is interesting to notice that the Ray Singer factor together with 
the $ZM$
factors dissapear for

\begin{equation}
p=\frac{d-2}{2},
\end{equation}

\noindent {}since $p$ is an integer $d$ is always even, and moreover, 
$p=%
\frac{d-2}{2}${} is the same relation which ensures that the coupling
constant has no dimensions. In general, however, the topological 
factors give
nontrivial contributions. Besides the evaluation of the topological 
factors
which appear in (\ref{eq:alv41}), the important point is that we have 
given
a precise definition of the space of local $p$-forms with nontrivial
transitions over $X$ which enter into the definition of the partition
function. Moreover, we have related it to the topological aspects
contained in the exact sequence of cohomology groups 
(\ref{eq:exact22}). In
particular we have identified them with the cohomology classes of 
$\check{H}%
^{p}(X,K^{P})$.

{\bf Acknowledgements:} This work was supported by the Decanato de
Investigaciones de la Universidad Simon Bolivar through the project 
DID-G11.


\begin{thebibliography}{9}

\bibitem{re:Weil} B. Kostant, Lecture Notes in Mathematics, Berlin, 
Springer (1970).

\bibitem{re:Brylinski}  J. L. Brylinsky, Prog. in Math. Vol 107, {\it 
Loop
Spaces Characteristic Classes and Geometric quantization}, Birkhauser 
Boston
(1993).

\bibitem{re:Murray}  A. L. Carey, M. K. Murray and B. L. Wang, 
hep-th/9511169

\bibitem{re:private}  J. L. Brylinsky and Mc. Laughlin (private
communication)

\bibitem{re:CMR}  M. I. Caicedo, I. Martin and A. Restuccia, 
hep-th/9701010.

\bibitem{re:Freed}  D. S. Freed, talk given at the XIX International
Colloquium on Group Theoretical Methods in Physics, Salamanca 1992.

\bibitem{re:Nepomichi}  R. Nepomechie, Phys. Rev. D31 (1985), 1921; C.
Teitelboim Phys. Lett {\bf B167} (1986) 69.

\bibitem{re:Steenrod}  S. Eilenberg and N. Steenrod, {\it Foundations 
of
Algebraic Topology}, Princeton, New Jersey, Princeton Univ. Press 
(1964).

\bibitem{re:BlauThom}  D. Birmingham, M. Blau, M. Rakowski and G. 
Thomson,
Phys. Rep. 209 (1991) 129-340.

\bibitem{re:Muller} W. M\"{u}ller, Adv.in Math.{\bf 28}(1978) 233.

\bibitem{re:Witten}  E. Witten, hep-th/9505186; E. Verlinde, Nucl. 
Phys. 
{\bf B455} (1995) 211; Y. Lozano, Phys. Lett. {\bf B364} (1995) 19; J. 
L. F.
Barb\'{o}n, Nucl. Phys. {\bf B452} (1995) 313; A. Kehagias, 
hep-th/9508159.
\end{thebibliography}
\end{document}